\documentclass[twocolumn]{aastex631}

\usepackage{graphicx}	
\graphicspath{ {figs} }
\usepackage{amsmath}	
\usepackage{amssymb}	
\usepackage{xcolor}
\usepackage{soul}
\usepackage{multirow}
\usepackage{enumitem}

\def\gizmo{\textsc{gizmo}}
\def\dice{\textsc{dice}}

\def\msol{\mathrm{M}_\odot}

\def\HI{\ion{H}{1}}

\def\CIV{\ion{C}{4}}

\def\SiIV{\ion{Si}{4}}

\def\OVI{\ion{O}{6}}
\def\Ha{H\,$\alpha$}

\newcommand{\rev}[2]{#2}
\newcommand{\revv}[2]{#2}

\defcitealias{lucchini20}{L20}
\defcitealias{lucchini21}{L21}

\begin{document}

\title{Properties of the Magellanic Corona}

\correspondingauthor{Scott Lucchini}
\email{scott.lucchini@cfa.harvard.edu}

\author[0000-0001-9982-0241]{Scott Lucchini}
\affiliation{Center for Astrophysics | Harvard \& Smithsonian, 60 Garden St, Cambridge, MA, USA}
\affiliation{Department of Physics, University of Wisconsin - Madison, Madison, WI, USA}

\author[0000-0003-2676-8344]{Elena D'Onghia}
\affiliation{Department of Physics, University of Wisconsin - Madison, Madison, WI, USA}
\affiliation{Department of Astronomy, University of Wisconsin - Madison, Madison, WI, USA}

\author[0000-0003-0724-4115]{Andrew J. Fox}
\affiliation{AURA for ESA, Space Telescope Science Institute, 3700 San Martin Drive, Baltimore, MD, USA}
\affiliation{Department of Physics \& Astronomy, Johns Hopkins University, 3400 N. Charles Street, Baltimore, MD 21218, USA}

\shortauthors{Lucchini et al.}



\begin{abstract} 

\rev{}{
We characterize the Magellanic Corona, the warm gaseous halo around the Large Magellanic Cloud (LMC). The Corona is a key ingredient in the formation of the Magellanic Stream (Lucchini et al. 2020, 2021) and has recently been observed in high-ion absorption around the LMC. In this work we present a suite of high-resolution hydrodynamical simulations to constrain its total mass and temperature prior to the infall of the Magellanic Clouds to our Galaxy. We find that the LMC is able to host a stable Corona before and during its approach to the MW through to the present day. With a Magellanic Corona of $>2\times10^9$ $\msol$ at $3\times10^5$ K, our simulations can reproduce the observed total mass of the neutral and ionized components of the Trailing Stream, size of the LMC disk, ionization fractions along the Stream, morphology of the neutral gas, and on-sky extent of the ionized gas. The Corona plays an integral role in the survival, morphology, and composition of the Magellanic Clouds and the Trailing Stream.}


\end{abstract}



\section{Introduction}

The Magellanic Stream is the largest coherent extragalactic gaseous structure in our sky \citep{mathewson74}. It has the potential to dramatically impact the future of the Milky Way (MW) by depositing billions of solar masses of gas into our circumgalactic medium (CGM) and possibly onto our disk \citep{fox14,donghia16}. The Magellanic Stream also provides direct evidence of galaxy interactions and evolution through mergers. By studying this serendipitous nearby system, we will learn about the future of our own Galaxy, the history of the Local Group, 
and the gas and metal transport processes that can sustain the growth of galaxies like the MW.

The Magellanic Stream is an extended network of interwoven clumpy filaments of gas that originate from within the Large and Small Magellanic Clouds (LMC, SMC), two satellite galaxies of the MW. Combined with the Leading Arm, high velocity clumps of gas ahead of the Magellanic Clouds in their orbits, the Magellanic System covers over 200$^\circ$ on the sky. From 21-cm \HI\ observations, we have a detailed view of its small-scale, turbulent morphology as well as its velocity structure \citep[e.g.][]{cohen82,morras83,putman03,bruns05,nidever08,nidever10,westmeier18}.
Moreover, absorption-line spectroscopy studies have characterized the chemical composition and ionization state of the Stream along dozens of sightlines \citep{lu94,gibson00,sembach03,fox10,fox13,fox14,richter13}.
\citet{fox14} has shown that the Stream is mostly ionized. They find an average ionization fraction of $\approx$73\% with a total ionized gas mass of $\sim1.5\times10^9$ $\msol$ (compared with $4.9\times10^8$ $\msol$ of neutral gas; \citealt{bruns05}).

Models of the formation of the Stream originally explained the stripped material as gas that was tidally pulled from the LMC through repeated interactions with the MW \citep{fujimoto76,davies77,lin77,lin82,gardiner96,connors06,diaz11}. This would result not only in stripped gas, but also in a tidally truncated dark matter halo. Thus, masses determined by rotation curve fitting \citep[$1.7\pm0.7 \times10^{10}$ $\msol$ within 8.7 kpc,][]{vandermarel14} would be sufficient for modeling the evolution of the Clouds and the formation of the Stream. However, updated proper-motion measurements of the LMC and SMC have shown that the Clouds are most likely on their first infall towards the Milky Way \citep{kallivayalil06,besla07,kallivayalil13}. A recent study has found a second passage orbit consistent with the observations, however further studies of the hydrodynamics are required to see if the Stream can be reproduced \citep{vasiliev24}. A first-infall scenario would require a higher LMC mass as it would not yet be tidally truncated.

\begin{figure}[!ht]
    \centering
    \includegraphics[width=\columnwidth]{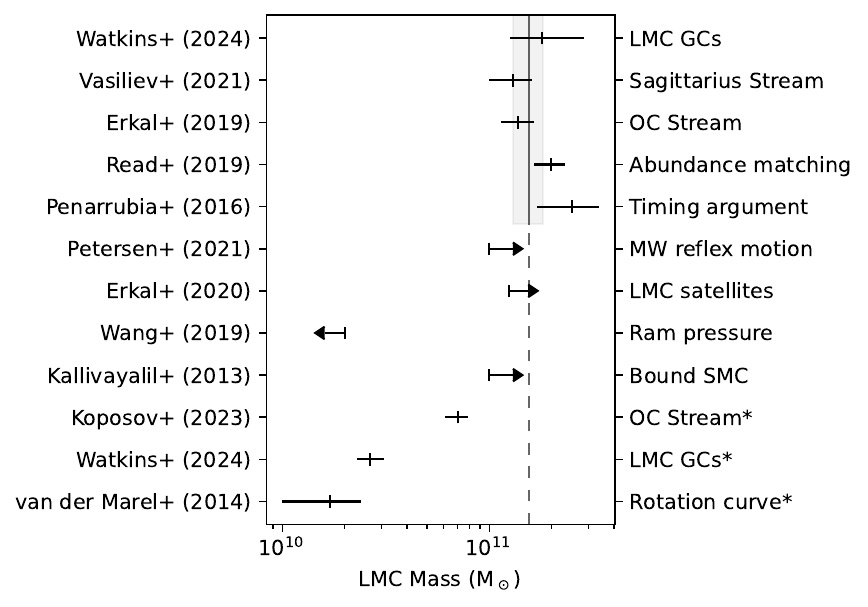}
    \caption{\textbf{Observational LMC total mass estimates}. Literature estimates for the total mass of the Large Magellanic Cloud. \citet{vandermarel14} and \citet{koposov23} (denoted with asterisks) provide constraints on the mass within 8.7 and 32.4~kpc, respectively, \rev{whereas all}{while \citet{watkins24} provides a direct constraint within 13.2~kpc (second from the bottom) in addition to a fit using an NFW profile (top). All} other estimates plotted are for the total virial LMC mass. \citet{kallivayalil13}, \citet{erkal20}, and \citet{petersen21} provide lower limits, and \citet{wang19} \revv{}{provide} an upper limit. The remaining references \citep{penarrubia16,read19,erkal19,vasiliev21,watkins24} give estimates of the total LMC mass with uncertainties. The error weighted average of these five values is \rev{$1.60\pm0.21\times10^{11}$}{$1.55\pm0.26\times10^{11}$ $\msol$} shown as the vertical line (extended to the bottom of the plot as a dashed line for comparison). The one-sigma errors are shown as a shaded region. The method for determining the LMC mass estimate from each paper is listed on the right edge of the plot.}
    \label{fig:lmcmasses}
\end{figure}

Many different indirect methods of estimating the LMC's total pre-infall mass have converged on a value of $1-2\times10^{11}$ $\msol$: 
$1.98\times10^{11}$ $\msol$ from abundance matching \citep{read19},
$>$10$^{11}$ $\msol$ from the MW's reflex motion \citep{petersen21},
$>$1.24$\times10^{11}$ $\msol$ from the LMC's satellite population \citep{erkal20}, $2.5^{+0.09}_{-0.08}\times10^{11}$ $\msol$ from the Hubble flow timing argument \citep{penarrubia16}, and \rev{$1-2\times10^{11}$ $\msol$ from the MW's stellar streams \mbox{\citep{erkal19,vasiliev21}}}{$1.3\pm0.3\times10^{11}$ and $1.38^{+0.27}_{-0.24}\times10^{11}$ $\msol$ from the Sagittarius \citep{vasiliev21} and Orphan-Chenab Streams \citep{erkal19}, respectively}.
See Figure~\ref{fig:lmcmasses} for a summary of these measurements. We include the error-weighted mean of the values calculated in \rev{}{\citet{watkins24},} \citet{vasiliev21}, \citet{erkal19}, \citet{read19}, and \citet{penarrubia16}: \rev{$1.6\pm0.21\times10^{11}$}{$1.55\pm0.26\times10^{11}$} $\msol$ ($\sim10-20$\% of the MW's total mass)\footnote{\rev{}{We note that some of these methods do depend on the assumed MW total mass, which is uncertain at the $\sim30$\% level \citep{bland-hawthorn16}. However, each of the references listed above have simultaneously constrained the MW and LMC mass to reach the values quoted. Moreover, even reducing the LMC mass estimates by 30\% leaves our average value above $10^{11}$ $\msol$.}}.

While modern tidal models of the formation of the Stream have used large masses for the LMC, they are unable to explain the immense mass of ionized gas \citep[$>10^9\msol$][]{besla10,besla12,pardy18}. On the other hand, ram pressure models \citep{meurer85,moore94,sofue94} are able to explain the ionized material via dissolution of the neutral gas through instabilities, but they require low masses for the LMC \citep[$<2\times10^{10}$ $\msol$;][]{hammer15,wang19}.

To resolve both of these discrepancies simultaneously, we introduced the Magellanic Corona model (\citealt{lucchini20,lucchini21}, hereafter \citetalias{lucchini20}, \citetalias{lucchini21}). Based on theoretical calculations and cosmological simulations, galaxies with masses $\sim10^{11}$ $\msol$ should host gaseous halos at or near their virial temperature of $\sim3\times10^{5}$ K\footnote{While this value is near the peak of the interstellar cooling curve, 
the Corona can remain stable if heated (e.g. through supernova feedback). See Figure~\ref{fig:coronagridtemp}.} \citep{jahn21}. 
Upon inclusion of a warm CGM around the LMC, dubbed the Magellanic Corona, we are able to explain the ionized material in the Stream while accounting for a massive LMC. In \citetalias{lucchini21}, we showed that this Magellanic Corona also exerts hydrodynamical drag on the SMC as it orbits around the LMC. With a new orbital history consistent with the present-day positions and velocities of the Clouds, the Trailing Stream ends up several times closer to us than previous models predicted. 

\begin{table*}
    \begin{center}
    \caption{Initial Galaxy Properties \rev{}{in the simulations}}
    \label{tab:ics2}
    \small
    \hspace{-1.4cm}
    \begin{tabular}{lccc}
        \hline
         & LMC ($t=0$ Gyr) & SMC & MW \\\hline
        DM Mass ($\msol$) & $1.75\times10^{11}$ & $5\times10^9$ & $1.1\times10^{12}$ \\
        Stellar Disk Mass ($\msol$) & $2.5\times10^9$ & $2.5\times10^8$ & -- \\
        Stellar Scale Length (kpc) & 4.5 & 1.7 & -- \\
        Gaseous Disk Mass ($\msol$) & 0 & $1.5\times10^9$ & -- \\
        Gaseous Scale Length (kpc) & -- & 6.9 & -- \\
        CGM Mass$^a$ ($\msol$) & $10^9-10^{10}$ & -- & $2\times10^{10}$\\
        CGM Temperature$^a$ (K) & $1-9\times10^5$ & -- & $10^6$ \\
        N Particles & $5.5\times10^5$ & $8.5\times10^4$ & $8.8\times10^5$ \\\hline
    \end{tabular}
    \end{center}
    \footnotesize{\raggedright \textbf{Notes.} These values are adopted in the simulations throughout this paper. They differ from \citetalias{lucchini21} in the stellar and gaseous disk masses for the Clouds (lowered to better match observed values), the MW only consists of a DM halo and its corona (for more efficient computation), and the total mass of the SMC has been decreased (to better facilitate the stripping of its ISM into the Trailing Stream).\\
    $^a$ The CGM mass and temperature for the LMC is shown as the range of values that we explore in Section~\ref{sec:corona}.}
\end{table*}

In a subsequent study, the Magellanic Corona was directly observed using absorption line spectroscopy data from the Cosmic Origins Spectrograph on the \textit{Hubble Space Telescope} \citep{dk22}. From 28 sightlines extending to 35 kpc away from the LMC, we detected a radially declining profile in high ions (\CIV, \SiIV, \OVI) with a total mass of $1.4\pm0.6\times10^9$ $\msol$ including a warm phase with temperature of $3\times10^{5}$ K. These results are consistent with the picture of \revv{a first-infall LMC with}{an LMC embedded within} a Magellanic Corona. While these values give us \revv{}{a detailed} picture of the LMC's CGM at the present-day, modeling the evolution of the Magellanic System is needed to constrain the properties of the primordial LMC and its Magellanic Corona. 

\rev{In this paper, we expand upon the Magellanic Corona model by presenting new, high-resolution simulations}{In this paper, we provide an in-depth exploration of the properties of the Magellanic Corona in the context of the formation of the Magellanic Stream. We present a suite of new simulations} with detailed physical models including the self-consistent tracking of star formation, feedback, ionization, and metallicity to directly compare with absorption line spectroscopy observations. \rev{Furthermore, w}{W}e explore the parameter space of temperatures and densities for the Magellanic and Galactic Coronae that provide the best match to the observed properties of the Magellanic Clouds as well as the morphology of the neutral \HI\ Stream.
In Section~\ref{sec:methods}, we outline the methodology of our simulations and analysis.
In Section~\ref{sec:corona}, we discuss our parameter space exploration of the properties of the Magellanic Corona.
Section~\ref{sec:discussion} contains the discussion of our results and the implications for the properties of the MW's CGM. We conclude in Section~\ref{sec:conclusions}.

\section{Methods} \label{sec:methods}

\rev{}{Our simulation methodology contains three distinct steps. First, we initialize an LMC-mass galaxy (the ``$t=0$ LMC''; Section~\ref{sec:ics}) and run it in isolation (Section~\ref{sec:coronastability}). Once that galaxy has come to equilibrium, we then set up a new simulation with the MW, SMC, and with that galaxy as our initial LMC (the ``pre-infall LMC''). Finally, we let them evolve to their present-day positions and compare with the observed properties of the Magellanic System (Section~\ref{sec:coronastream}). Based on these comparisons with observations, we can determine the best values for our initial LMC and Magellanic Corona (our ``fiducial'' mass and temperature).}

\subsection{Simulation Setup} \label{sec:sim}

We use \gizmo, a massively parallel, multiphysics code for our simulations \citep{hopkins15,springel05}. We utilize its Lagrangian ``meshless finite-mass'' hydrodynamics scheme which allows for the ability to track individual fluid elements while conserving angular momentum and capturing shocks \citep{hopkins15}.
Star formation is included \citep{springel03} with a physical density threshold of 100 cm$^{-3}$, a virial requirement that the gas is locally self-gravitating \citep{hopkins13,hopkins18}, and a requirement that the gas is converging ($\nabla\cdot v<0$). Mechanical stellar feedback is also included in which we assume a constant supernova rate of $3\times10^{-4}$ SNe Myr$^{-1}$ $\msol^{-1}$ for all stars less than 30 Myr old. Each supernova injects 14.8 $\msol$ with $10^{51}$ ergs of energy and metals following the AGORA model \citep{kim16}. Cooling is included down to $\sim10$ K following \citet{hopkins18} with metal-dependent tables \citep{wiersma09}. We don't include radiative transfer or UV background radiation, however outside the MW at low redshift, we don't expect these mechanisms to play a significant role.

These simulations improve upon those of \citetalias{lucchini20} and \citetalias{lucchini21} by including accurate star formation and feedback with metallicity and advanced cooling routines. \gizmo\ calculates self-consistent ionization states for each particle based on collisional heating/ionization, recombination, free-free emission, high and low temperature metal-line cooling, and compton heating/cooling \citep[see Appendix~B in][]{hopkins18} We therefore are able to track the neutral and ionized material separately throughout the simulation.


\subsection{Initial Conditions} \label{sec:ics}

Table~\ref{tab:ics2} shows the properties of the Magellanic Clouds in our simulation and we used the \dice\footnote{\url{https://bitbucket.org/vperret/dice/src/master/}} code to generate our initial conditions \citep{dice}.
We used an LMC dark matter (DM) mass of $1.75\times10^{11}$ $\msol$ consistent with previous studies (\citealt{besla12,pardy18}; \citetalias{lucchini20,lucchini21}) and in agreement with indirect estimations (see Figure~\ref{fig:lmcmasses}). We constrained the initial gaseous and stellar disk masses for the Magellanic Clouds by requiring their present-day values to be consistent with observations. This was straightforward for the stellar masses as the stars formed during the simulations comprise only a small fraction of the total. \rev{}{Our $t=0$ LMC includes a stellar disk of $2.5\times10^9$ $\msol$.} However the gas masses can vary greatly from their initial values due to the tidal interactions and accretion from the Magellanic Corona.

\rev{We found that the LMC disk gas mass agreed best with present-day values when allowing the gaseous disk to form naturally out of the Magellanic Corona.}{
After initializing our Magellanic Corona in \dice\ with a fixed temperature, we find that it takes $\sim4$~Gyr for the system to settle into equilibrium. During this time, 18\% of the Coronal gas (in the fiducial case, see below) cools and falls to the center of the potential which increases the gas mass in the galactic disk. This condensation of the Corona forms enough of a gas disk that, at the present day, it is consistent with observations of the LMC. Therefore, we initialize our $t=0$ LMC with a DM halo, stellar disk, and Magellanic Corona. And after $\sim4$ Gyr of evolution in isolation, we find a stable LMC galaxy that we use to represent out ``pre-infall'' LMC.
}
\rev{Therefore, we initialize our fiducial LMC ($t=0$ Gyr in Table~\mbox{\ref{tab:ics2}}) with a $2.5\times10^9$ $\msol$ stellar disk and a Magellanic Corona with total mass (within 200 kpc) of $5\times10^9$ $\msol$}{}{}

\begin{figure}
    \centering
    \includegraphics[width=\columnwidth]{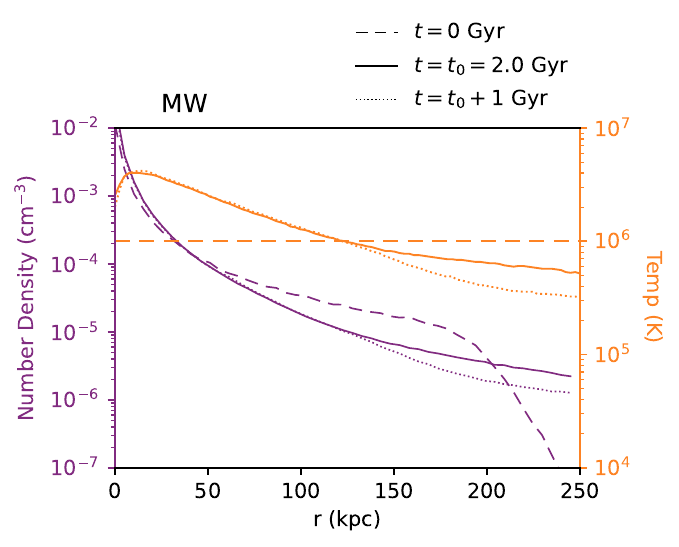}
    \caption{\textbf{Initial properties of the Galactic Corona}. The number density and mean temperature are shown as a function of radius in black and orange, respectively. The dashed lines \revv{}{show} the profiles at $t=0$ Gyr, and the solid lines show the profiles at the end of isolated evolution ($t_0=2$ Gyr). The dotted lines show the profiles after an additional 1 Gyr of evolution.}
    \label{fig:coronaic}
\end{figure}

\rev{}{In Section~\ref{sec:corona} below, we explore a variety of masses and temperatures for the Magellanic Corona. In our fiducial case, we initialize our LMC with a Magellanic Corona containing $5\times10^9$ $\msol$} following an isothermal profile \rev{}{(exponentially truncated at 200~kpc)} at a temperature of $3\times10^5$ K and a metallicity of 0.1 solar. After 3.5 Gyr in isolation, there remains $3.7\times10^9$ $\msol$ of ionized material bound to the LMC (with $2\times10^9$ $\msol$ within 120~kpc) with a median temperature of $3.4\times10^5$ K. $9.1\times10^8$ $\msol$ has cooled to form the LMC's gaseous disk. \rev{This is the LMC that we start with for the full simulations with the LMC, SMC, and MW (listed as ``LMC ($t=3.5$ Gyr)'' in Table~\mbox{\ref{tab:ics2}}).}{We adopt this as our fiducial ``pre-infall" LMC.} \rev{The initial and final r}{R}adial profiles of density and temperature are shown in Figure~\ref{fig:coronaic} in blue.\rev{ A parameter space exploration of optimal masses and temperatures for the Magellanic Corona is discussed in Section~\mbox{\ref{sec:corona}}.}{}

The only constraint on the SMC's total mass comes from its rotation curve, which requires $2.4\pm0.36\times10^9$ $\msol$ within 4 kpc ($1.25\pm0.25\times10^9$ $\msol$ in DM, \citealt{diteodoro19}; \citealt{stanimirovic04}). \rev{}{However, recent work has shown that the SMC is possibly much more complicated than originally thought \citep{murray23}.} \rev{While previous studies found that $\sim10\%$ of the LMC's total mass ($\sim2\times10^{10}$ $\msol$) produced the best results \mbox{\citep{pardy18}}, this was prior to the inclusion of the Magellanic and Galactic Coronae. Therefore, in this study we have explored the effect of the SMC total mass on the formation of the Stream and found that a lower SMC mass provides better results.}{A variety of SMC masses have been used in previous models ($10^9-3\times10^{10}$ $\msol$; \citealt{besla12,diaz12,pardy18,zivick18,deleo20,patel20,cullinane22}) and with the inclusion of the Magellanic and Galactic Coronae, we found that lower SMC masses allow for stripped material in agreement with the observations when accounting for realistic heating, cooling, and ionization.}
In our fiducial model, we used a DM mass of $5\times10^9$ $\msol$, a stellar mass of $2.5\times10^8$ $\msol$, and a gaseous disk mass of $1.5\times10^9$ $\msol$ (listed in Table~\ref{tab:ics2}).

For the MW, we implemented a live DM halo with a total mass of $1.1\times10^{12}$ $\msol$ combined with a hot gaseous CGM following a $\beta$-profile as in \citet{salem15}:
\begin{equation}
    \rho \propto \left[1+ \left(\frac{r}{r_c}\right)^2\right]^{-3\beta/2}
\end{equation}
with $r_c = 0.35$ and $\beta=0.559$. As with the Magellanic Corona, we explored a variety of models for the Galactic corona as well, which will be discussed in Section~\ref{sec:mwcgm}. Our fiducial MW CGM has a total mass of $2\times10^{10}$ $\msol$ at $10^6$ K and is nonrotating. After evolution in isolation for 2 Gyr, $1.9\times10^{10}$ $\msol$ remains bound with a mean temperature of $1.4\times10^6$ K. Figure~\ref{fig:coronaic} shows the initial and final (after 2 Gyr in isolation) density and temperature profiles in orange.

We use a mass resolution of $\sim3\times10^4$ $\msol$ per particle for the gas elements, $\sim2\times10^4$ $\msol$ per particle for the stars, $\sim10^6$ $\msol$/particle for the DM. This results in a total particle number of $1.5\times10^6$. Adaptive softening was used for the gas particles (such that the hydrodynamic smoothing lengths are the same as the gravitational softening length), and softening lengths of 150 pc and 550 pc were used for the stellar and dark matter particles, respectively.

\subsection{Orbits of the Clouds}

\revv{}{
For our simulations including the full interactions between the Clouds and the MW, we used orbital histories very similar to those previously published in \citetalias{lucchini21} including two interactions over $\sim4$ Gyr. Our initial positions and velocities were
\begin{align*}
    r_{i,\mathrm{LMC}}&=(33.03, 584.97, 300.80)\\
    v_{i,\mathrm{LMC}}&=(2.24, -63.93, -65.77)\\
    r_{i,\mathrm{SMC}}&=(-21.40, 610.74, 357.15)\\
    v_{i,\mathrm{SMC}}&=(6.05, 10.48, -109.69)
\end{align*}
while the MW was initially at the origin with zero velocity.
}

\section{Properties of the Magellanic Corona} \label{sec:corona}

We explored the parameter space of initial properties of the Magellanic Corona, by varying the initial temperature and total mass. The Corona was initialized with an isothermal distribution. 
Its total mass (within 200 kpc) ranged from $10^9$ to $10^{10}$ $\msol$. This corresponds to masses of 0.6 and $6\times10^9$ $\msol$ within the LMC's virial radius of 120 kpc. The initial temperature of the Corona ranged from $10^5$ to $9\times10^5$ K, and we used a metallicity of 0.1 solar. We explored the viability of these different Coronae by determining their stability and their impact on the present-day Stream.

As mentioned above, we initialize our LMC with a DM halo, stellar disk, and the Magellanic Corona. The gaseous disk forms during the first few billion years of evolution (in isolation). The Magellanic Corona is initialized with a streaming fraction of 0.2, meaning it has an azimuthal velocity set to 20\% the circular velocity profile. Therefore, when the cooled material collapses onto the disk, it exhibits a bulk rotation as expected. Higher \rev{}{(lower)} streaming fractions result in larger \rev{}{(smaller)} disks.\rev{and lower streaming fractions result in smaller disks.}{} This is because without any rotation, more material falls into the center of the gravitational potential and high gas densities induce very strong star formation which blows out the remaining gas. With too much rotation, the infalling cool material spreads out to larger radii (because it has higher angular momentum) and the densities are not high enough for star formation.

\begin{table*}
    \caption{\rev{}{Pre-infall Properties of the LMC}}
    \label{tab:simsetups}
    \begin{center}
    \small
    \hspace{-2.3cm}
    \begin{tabular}{lccccccc}
        \hline
        Model & $t_0$ & \multicolumn{2}{c}{Ionized gas mass} & \multicolumn{2}{c}{Gaseous disk} & \multicolumn{2}{c}{Stellar disk} \\
         & (Gyr) & Bound & $r<120$~kpc & Mass & Scale length & Mass & Scale length \\
         &  & ($10^8$ $\msol$) & ($10^8$ $\msol$) & ($10^8$ $\msol$) & (kpc)  & ($10^8$ $\msol$) & (kpc)\\\hline
        Fiducial & 3.5 & 37.2 & 20.2 & 9.1 & 2.8 & 26.7 & 4.2 \\
        Low mass & 4.0 & 9.4 & 5.2 & 0.7 & 1.7 & 25.1 & 4.2 \\
        High mass & 2.5 & 88.9 & 36.3 & 6.1 & 2.2 & 30.1 & 3.7 \\
        Low temp & 4.0 & 29.6 & 21.7 & 13.6 & 23.5 & 29.6 & 3.8 \\
        High temp & 4.0 & 47.0 & 14.5 & 0.02 & -- & 26.3 & 4.0 \\\hline
    \end{tabular}
    \end{center}
    \footnotesize{\raggedright \textbf{Notes.} Column 1 lists the name of the model (corresponding to the $x-$axis of Figure~\ref{fig:coronagrid}), Column 2 lists the amount of time the LMC was evolved in isolation, Column 3 lists the amount of bound ionized gas mass (no radius cut), Column 4 lists the ionized gas within the LMC's virial radius (120~kpc), Columns $5-8$ list the masses and scale lengths of the gaseous and stellar disks. These are the properties of the LMC before its interactions with the SMC and approach towards the MW.}
\end{table*}

\subsection{Stability} \label{sec:coronastability}

The main factor in determining the viable parameter space for the temperature and density of the Magellanic Corona is its stability. If the temperature is too high, the Coronal plasma becomes unbound from the LMC and blows off into the Local Group. If the temperature is too low, too much gas falls onto the LMC, leading to disk gas fractions and star-formation rates that are too high compared to observations. Similarly, if the Corona starts with too much mass (i.e. too high density), the LMC disk becomes too gas-rich. Below a certain mass threshold, the Corona remains stable, but in order to reproduce the high ionization fractions along the length of the Stream, the Magellanic Corona must be more massive than $10^9$ $\msol$ (within 120 kpc; see Section~\ref{sec:coronastream}).

\begin{figure*}[!ht]
    \centering
    \includegraphics[width=0.85\textwidth]{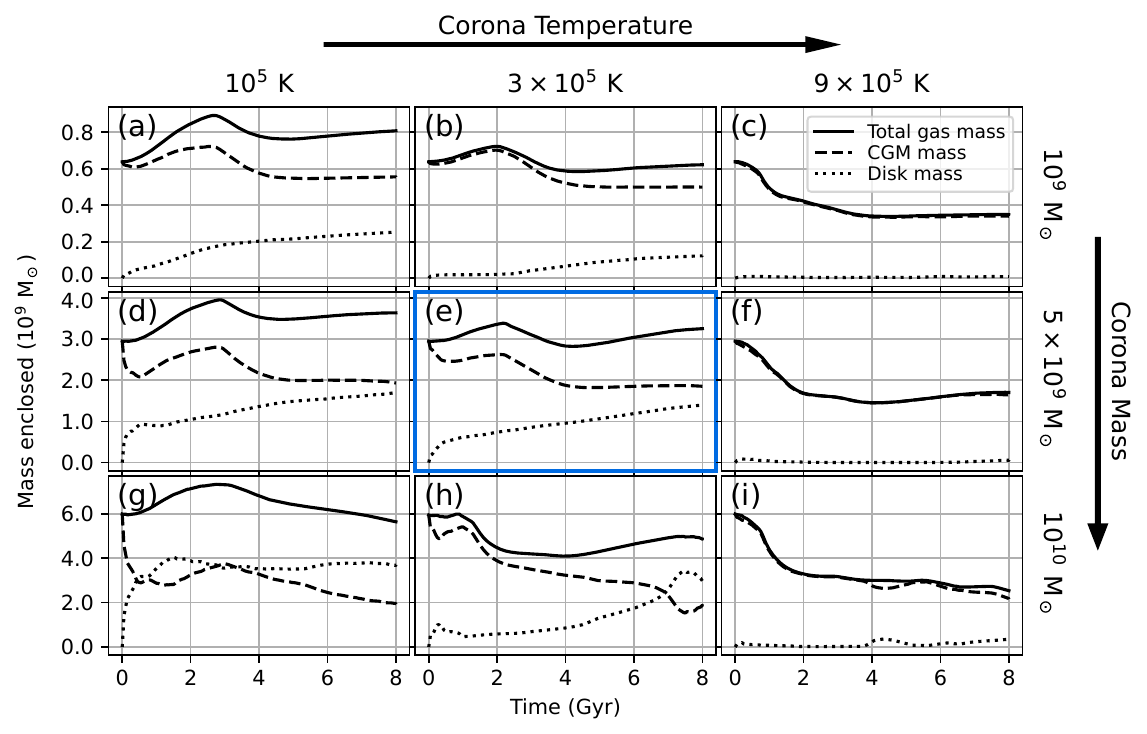}
    \caption{\textbf{Magellanic Corona stability}. Study of the stability of the Magellanic Corona for varying initial masses and temperatures. Each panel shows the total, CGM, and disk masses within 120 kpc as a function of time (solid, dashed, and dotted lines, respectively). 
    The initial temperatures increase from left to right with values of 1, 3, and $9\times10^5$ K. The initial masses increase from top to bottom with values of 1, 5, and $10\times10^9$ $\msol$ (within 200 kpc). Our fiducial model \rev{(see Section~\mbox{\ref{sec:fiducialcorona}})}{(outlined in blue)} is the center frame (Panel e) corresponding to an initial mass of $3\times10^9$ $\msol$ and a temperature of $3\times10^5$ K. \rev{}{The low and high mass models referenced in Figure~\ref{fig:newbars} correspond to Panels b and h, and the low and high temp models referenced in Figure~\ref{fig:lmcdisk} correspond to Panels d and f.}}
    \label{fig:coronagrid}
    \includegraphics[width=0.85\textwidth]{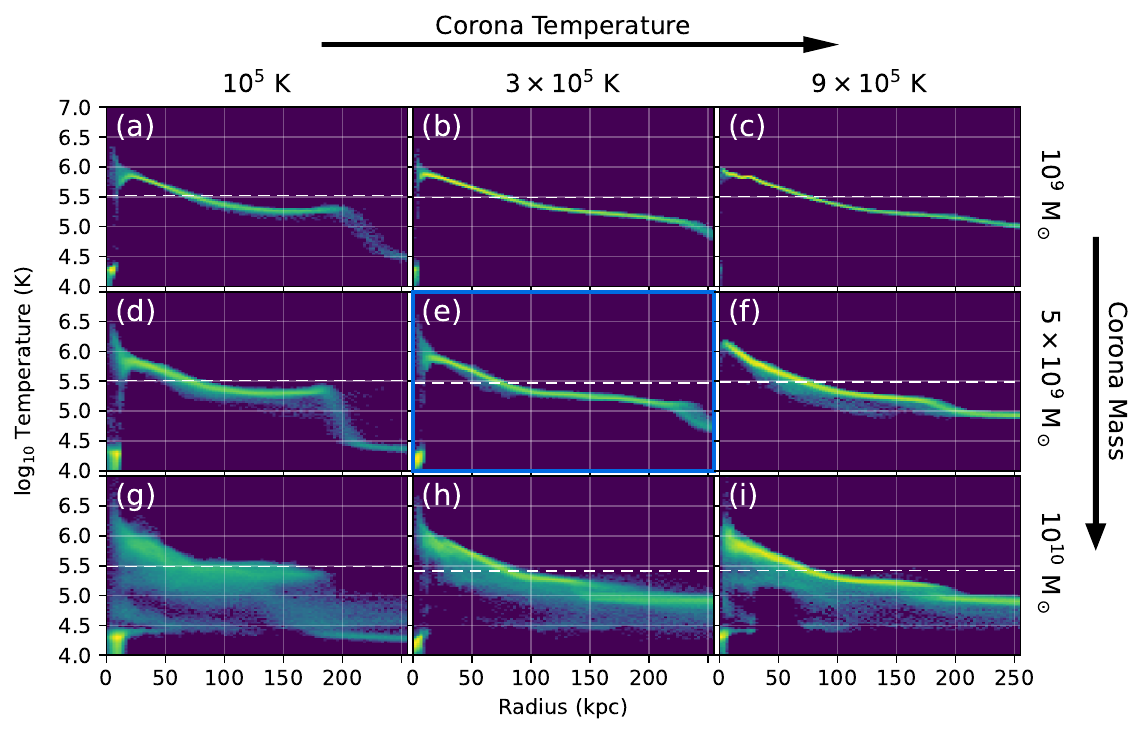}
    \caption{\textbf{Magellanic Corona temperature}. Temperature profile of the Magellanic Corona for nine initial masses and temperatures (as in Figure~\ref{fig:coronagrid}). These histograms show the relative density of the gas temperature as a function of radius for the nine simulations at $t=4$ Gyr. The horizontal white dashed lines show the mean temperature of the gas within $20<r<250$ kpc. \rev{}{Again, our fiducial model is Panel e, outlined in blue.}}
    \label{fig:coronagridtemp}
\end{figure*}

Figure~\ref{fig:coronagrid} shows these results. The nine panels depict nine different simulations with varying initial conditions in which the LMC and Magellanic Corona were evolved in isolation. The initial temperature of the Corona increases from left to right (with values of 1, 3, and $9\times10^5$ K), while the initial mass of the Corona increases from top to bottom (1, 5, and $10\times10^9$ $\msol$ within 200 kpc). The black lines show the total masses of the gaseous components within 120 kpc (the virial radius of the LMC) as a function of time -- total gas mass (solid), ionized mass (circumgalactic Coronal gas; dashed), and neutral disk mass (dotted).

Figure~\ref{fig:coronagridtemp} also shows the temperature distribution as a function of radius for the nine simulations at $t=4$ Gyr. Initial temperature increases to the right and the initial gas mass increases downwards. The mean temperature of the gas within $20<r<250$ kpc is shown as a horizontal dashed line. Interestingly, these white lines do not vary dramatically between the three columns. This means that the initial temperature has a minimal effect on the final stable temperature of the Corona.

We do, however, see that increasing the initial mass of the Corona strongly affects the range of temperatures. This is likely related to gas density; the Coronae with higher initial masses contain higher gas densities, which cools more effectively. Cooling is very efficient around $10^5$ K, so subtle changes in density and temperature can have a strong impact on the strength of cooling. These higher-density Coronae do not have sufficient supernova energy injection to keep the gas warm, and therefore result in lower temperatures.


\subsection{Effect on the Magellanic Stream} \label{sec:coronastream}

We now explore the effect of varying the Magellanic Corona's initial mass and temperature on the properties of the present-day Magellanic System.

\subsubsection{Mass} \label{sec:coronastream-mass}

\begin{figure}[!ht]
    \centering
    \includegraphics[width=0.9\columnwidth]{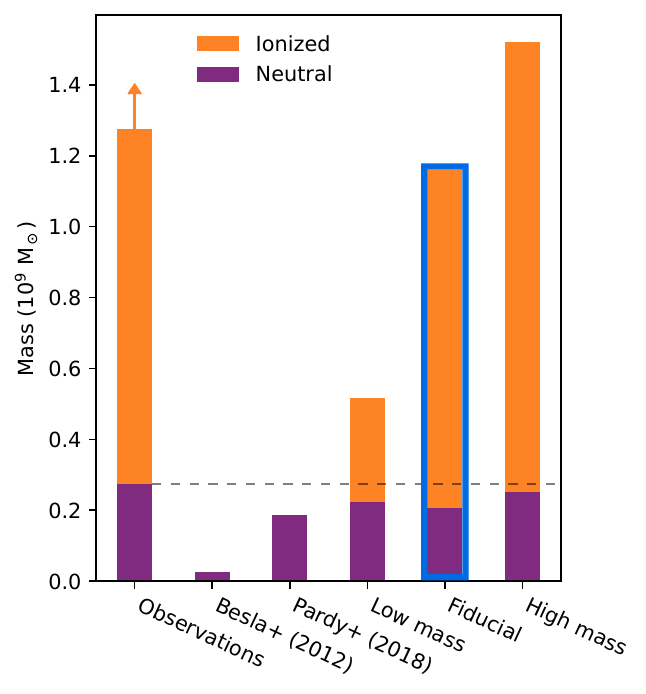}
    \caption{\textbf{The mass of the Trailing Stream}. Each bar shows the results for a different model compared against the observed mass in neutral and ionized gas shown in the left-most column ($2.7\times10^8$ $\msol$ in neutral gas, \citealt{bruns05}; $1.0\times10^9$ $\msol$ in ionized gas, though this may be an underestimate, \citealt{fox14}). Each value is computed by integrating the column densities for all gas behind the SMC ($l_\mathrm{MS} < l_\mathrm{MS, SMC}$) assuming a distance to the gas of 55 kpc (as done in observational works, \citealt{bruns05}, \citealt{fox14}). The total neutral masses in \citet{besla12}, and \citet{pardy18} are $2.5\times10^7$ $\msol$ and $1.9\times10^8$ $\msol$, respectively. The three right-most columns show the results from our new simulations with \rev{initial Magellanic Corona masses of 1, 5, and $10\times10^9$ $\msol$ (at a temperature of $3\times10^5$ K}{our low mass, fiducial, and high mass LMCs (each with initial temperature as in the fiducial model} ; Panels b, e, and h in Figures~\ref{fig:coronagrid} and \ref{fig:coronagridtemp}\rev{}{; see also Table~\ref{tab:simsetups}}). \rev{}{Our fiducial model is outlined in blue (as done in Figures~\ref{fig:coronagrid} and \ref{fig:coronagridtemp}).} The neutral material is relatively consistent between these models at 2.2, 2.1, and $2.5\times10^8$ $\msol$, while the ionized gas masses are $3.0\times10^8$, $9.5\times10^8$, and $1.3\times10^9$ $\msol$ for the \rev{low, medium, and high mass}{three} models, respectively. By dividing the ionized mass by the total mass, we can get an approximate value for the average ionization fraction in the Magellanic Trailing Stream, and we find values of 57\%, 82\%, and 84\% for the three models (compared with $\sim75\%$ in the observed case).}
    \label{fig:newbars}
\end{figure}

\begin{figure*}[!ht]
    \centering
    \includegraphics[width=0.8\textwidth]{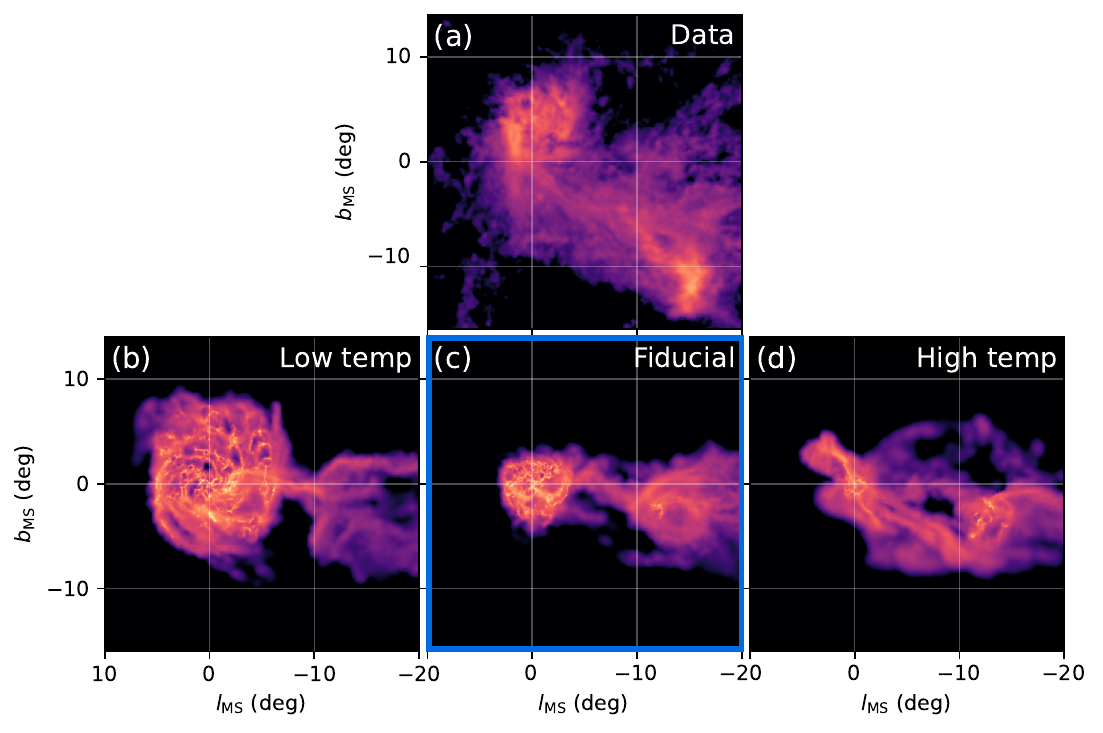}
    \caption{\textbf{Present-day LMC disk for different Corona temperatures}. A zoomed in region around the Magellanic Clouds is shown in Magellanic Coordinates for the observational data (Panel a; HI4PI, \citealt{westmeier18}) compared with three different simulations (bottom panels). Panels b, c, and d show the results from \rev{initial Magellanic Corona temperatures of 1, 3, and $9\times10^5$ K}{our simulations with low temp, fiducial, and high temp LMCs}, respectively (\rev{with a total mass of $5\times10^9$ $\msol$}{each with its total mass set to the fiducial value}; Panels d, e, and f in Figures~\ref{fig:coronagrid} and \ref{fig:coronagridtemp}\rev{}{; see also Table~\ref{tab:simsetups}}). \rev{}{Our fiducial model is outlined in blue (as done in Figures~\ref{fig:coronagrid} and \ref{fig:coronagridtemp}).} As the temperature decreases the size of the LMC disk increases due to Corona material collapsing onto the disk. At higher temperatures (Panel d), barely any LMC disk forms.}
    \label{fig:lmcdisk}
\end{figure*}

The initial total mass of the Corona directly affects the amount of ionized gas that we observe around the Stream today. Using UV absorption-line spectroscopy and photoionization modeling, \citet{fox14} estimated that there is $\sim1.5\times10^9$ $\msol$ of ionized gas associated with the Magellanic System, $\sim10^9$ $\msol$ of which is in the Trailing Stream region. Figure~\ref{fig:newbars} shows the total mass in the Trailing Stream for various different models. These values were calculated by mimicking the observational technique of integrating the column density at an assumed distance of 55 kpc\footnote{$M=m_p\Delta x \Delta y \sum N$, where $m_p$ is the proton mass, $\Delta x$ and $\Delta y$ are the physical sizes (in cm) of the bin widths in our column density image ($\Delta x=\Delta l_\mathrm{MS} \times D$, where $D$ is the assumed distance), and $N$ is the column density.}. These are not the physical masses in the system\rev{(see Section~\mbox{\ref{sec:physicalmasses}})}{}, but allow us to compare directly with the observational estimates shown in the left-most bar \citep{bruns05,fox14}. Continuing from left to right we have the results from the simulations published in \citet{besla12} and \citet{pardy18} in which no ionized material was produced. The three right-most bars show the results of our simulations for three different initial Corona masses\rev{, 1, 5, and $10\times10^9$ $\msol$ (corresponding}{}.

\rev{}{The initial LMCs used in these simulations are the ``low mass,'' ``fiducial,'' and ``high mass'' models corresponding} to Panels b, e, and h in Figures~{\ref{fig:coronagrid} and \ref{fig:coronagridtemp}. \rev{}{The properties of these galaxies after evolving in isolation are listed in Table~\ref{tab:simsetups}. We take these LMCs and place them at their initial positions with the SMC and MW and allow them to evolve until they reach their present-day positions. From these simulations, we calculate the total neutral and ionized material in the Trailing Stream; these values correspond to the three right-most columns in Figure~\ref{fig:newbars}.}
Clearly, as we increase the progenitor mass, we reproduce more ionized material in the Stream. For masses below $5\times10^9$ $\msol$, we are unable to reproduce the observations. In the models that we tested, either \rev{$5\times10^9$ or $10^{10}$ $\msol$}{the fiducial or high mass models} result in viable \rev{models}{ionized gas masses}. For \rev{}{higher} masses, \rev{larger than $10^{10}$ $\msol$, the mass of}{} the LMC's CGM begins to approach estimates of the MW CGM mass, which is unrealistic given the significant difference in virial masses (a factor of $\approx$10) of the two galaxies. Therefore Magellanic Corona masses below \rev{$10^{10}$ $\msol$}{our high mass model value ($\sim10^{10}$ $\msol$)} are preferred.

\subsubsection{Temperature} \label{sec:coronastream-temp}

We explore the impact of the initial Magellanic Corona temperature on the present-day Magellanic System. As shown in Figure~\ref{fig:coronagridtemp}, the initial temperature does not have a large effect on the temperature distribution or on the mean Corona temperature after 4 Gyr of evolution. The main difference that we see between temperature models is in the properties of the LMC disk. \rev{Since we form our LMC disk self-consistently by letting it condense out of the Corona}{Because the initial LMC gas disk is formed out of the condensation of the Corona in our isolated equilibrium simulations}, the temperature plays a large role in its size and stability.

\rev{}{As above (Section~\ref{sec:coronastream-mass}), we have taken our pre-infall initial LMC galaxies and run full simulations of the infall of the Clouds towards the MW. In this case, we are using the ``Low temp,'' ``Fiducial,'' and ``High temp'' models corresponding to Panels d, e, and f in Figures~\ref{fig:coronagrid} and \ref{fig:coronagridtemp}. Their properties are again listed in Table~\ref{tab:simsetups}.}

Figure~\ref{fig:lmcdisk} shows the LMC's disk at the present day for \rev{}{these} three \rev{different}{} simulations compared with observational data from the HI4PI survey \citep{hi4pi,westmeier18}. Panels a, b, and c show the results from \rev{simulations with initial Magellanic Corona Masses of 1, 3, and $9\times10^5$ K}{the low temp, fiducial, and high temp simulations}, respectively. Lower initial Corona temperatures lead to larger LMC disks due to more material cooling and falling towards the center of the gravitational potential. For the highest temperatures (Panel c), no LMC disk forms since the Corona material can't cool and fall to the center of the gravitational potential. This is also visible in Figure~\ref{fig:coronagrid}f in which the dotted line (showing the total disk mass) remains at zero throughout the simulation \rev{}{and in Table~\ref{tab:simsetups} in which the total neutral disk mass is shown to be $2\times10^6$ $\msol$}.

\section{Discussion} \label{sec:discussion}

\rev{}{The Magellanic Corona model of the evolution of the Magellanic System \citepalias{lucchini20, lucchini21} has been shown to be a viable candidate for the explanation of the high ionized mass observed \citep{fox14}. Strong support for the Corona model is given by the LMC's high total mass ($>10^{11}$ $\msol$), since a massive LMC can sustain a warm-hot CGM \citep{jahn21}. In this work we have explored the physical properties of the Magellanic Corona. We find it has pre-infall masses $\gtrsim2\times10^{9}$ $\msol$ within the virial radius (Table~\ref{tab:simsetups}) and gas with temperatures ranging from $10^5-10^6$ K (Figure~\ref{fig:coronagridtemp}).}

\rev{}{These results are consistent with previous works quantifying the circumgalactic gass of LMC-mass galaxies in cosmological simulations. \citet{jahn21} found circumgalactic gas masses of $3-6\times10^9$ $\msol$ at temperatures of $10^{4.5}-10^6$ K for halos with $M_{200}\sim1.5\times10^{11}$ $\msol$ in the FIRE simulations \citep{hopkins18}, in direct agreement with our findings. Furthermore, \citet{hafen19} characterized the CGM of galaxies across a range of halo masses also using the FIRE simulations. For their $10^{11}$ $\msol$ halos, they found an average CGM mass of $0.2f_bM_{200}\sim3\times10^9$ $\msol$.}

Recently, \citet{setton23} explored the implications of the bow shock generated as the LMC approaches the MW. The envelope of shocked gas could possibly explain some of the ionized gas detections of \citet{fox14}. Our model presented here does produce some of the same bow shock features including a high temperature leading edge of ionized gas. Because of the inclusion of the Magellanic Corona in our models, there is less of a clear shock boundary, however we are still consistent with the observed H$\alpha$ emission and ionized gas properties.

\subsection{Constraining the Milky Way Corona} \label{sec:mwcgm}

By exploring the parameter space of temperatures and densities for the MW's own hot gas corona, we can isolate its effects on the formation and morphology of the Trailing Stream. In this way, we can constrain the CGM properties by comparing our simulations with observations. Inspired by observations \citep[e.g.][]{anderson10,bregman18}, we varied the total mass from $10^{10}$ to $8\times10^{10}$ $\msol$, varied the temperature from $4\times10^5$ to $3\times10^6$ K, and tested with and without uniform rotation of the CGM. As we found above with the Magellanic Corona, changing the initial temperature of the gas did not affect the equilibrium temperature distribution significantly. Similarly, rotation, while it did decrease the equilibrium temperature of the CGM slightly, did not have a substantial effect on the Magellanic System. Therefore, the main variable we explored was the total mass.

\citetalias{lucchini21} found that a MW corona mass of $4\times10^{10}$ $\msol$ was required to get the best match to the velocity gradient along the Stream (two times larger than \citealt{salem15}). In the suite of orbits tested in this paper, we found that the largest effect that the MW corona had on the present-day Stream is on the morphology of the neutral and ionized components. Figure~\ref{fig:mw_cgm_effect} shows the ionized (orange) and neutral (grayscale contours) components of the simulated Streams in Magellanic Coordinates for three different models. The total mass of the MW corona increases from top to bottom with values of 1, 2, and $8\times10^{10}$ $\msol$. The higher gas density induces stronger ram pressure on the Magellanic gas, decreasing the on-sky extent of the ionized gas, and making the neutral Stream longer and narrower. We find that a value of $2\times10^{10}$ $\msol$ (Panel b; in agreement with estimates from \citealt{salem15}) provides the best agreement with the observations.

This led to our fiducial MW corona model with a total mass of $2\times10^{10}$ $\msol$ and a temperature of $10^6$ K. After 2 Gyr of evolution in isolation, $1.9\times10^{10}$ $\msol$ remains bound to the MW and the coronal gas has a mean temperature of $1.4\times10^6$ K.

\begin{figure}
    \centering
    \includegraphics[width=\columnwidth]{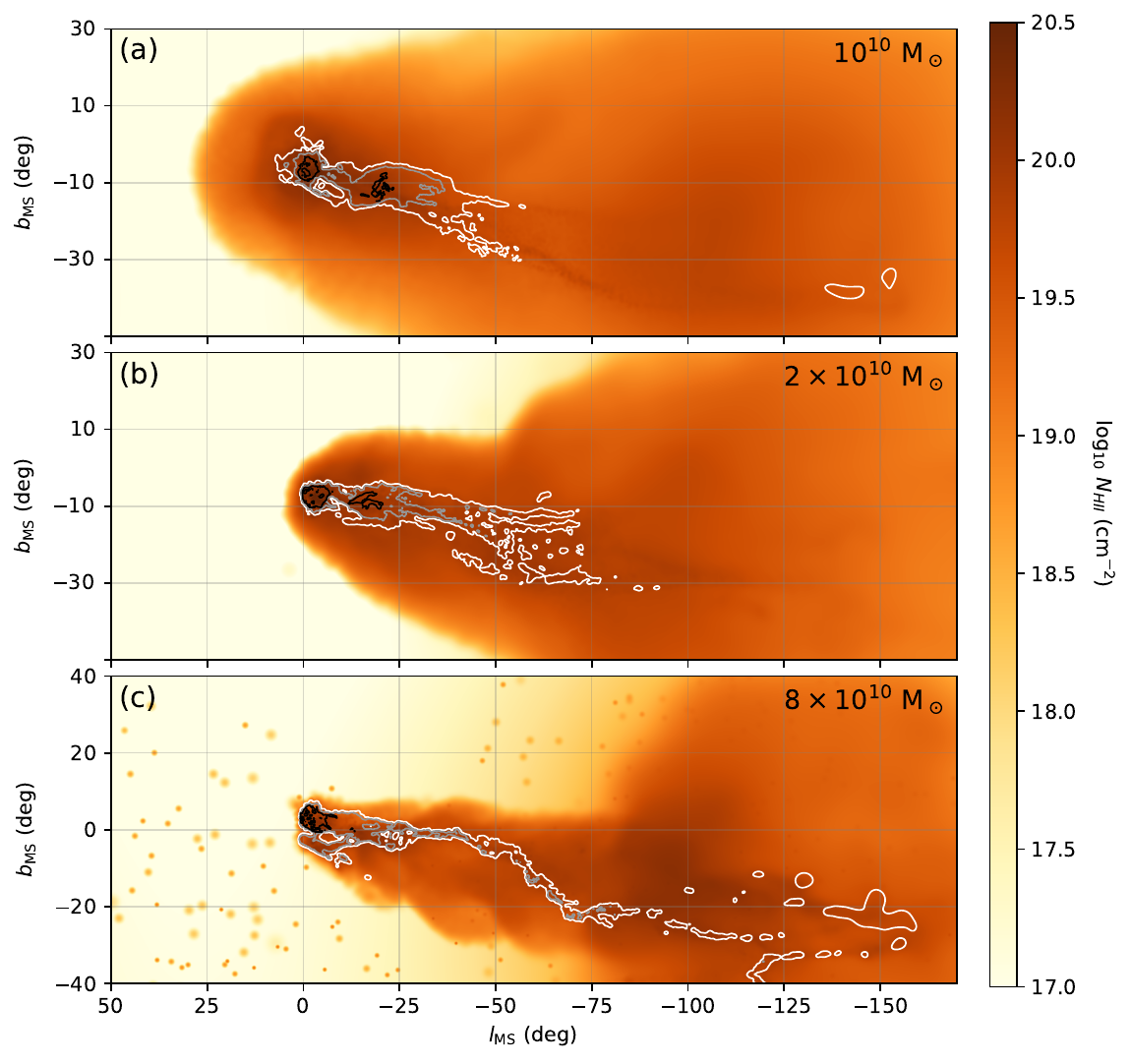}
    \caption{\textbf{The Effect of the MW Corona on the Present-Day Stream}. The ionized (orange) and neutral (grayscale contours) Magellanic gas for three simulations shown as they would appear on the sky in Magellanic Coordinates. Panels a, b, and c, show the results of simulations with a total MW coronal mass of 1, 2, and $8\times10^{10}$ $\msol$, respectively (listed in the top right of each panel). All panels use our fiducial LMC with a Magellanic Corona of $5\times10^9$ $\msol$ at $3\times10^5$ K. The neutral gas contours (white, grey, and black) are at values of log$_{10}(N_\mathrm{HI})=19$, 20, and 21, respectively.}
    \label{fig:mw_cgm_effect}
\end{figure}

\section{Conclusions} \label{sec:conclusions}

Building on our earlier work \citepalias{lucchini20,lucchini21}, we have characterized the influence of the Magellanic Corona on the formation and evolution of the Magellanic Stream.
With this suite of simulations, we have shown that the first-infall Magellanic Corona model of Stream formation can produce a present-day Magellanic System with \rev{}{global} properties in agreement with the observations. \rev{The ionized component is formed out of the Magellanic Corona, which becomes warped and shaped around the Clouds and the neutral Stream through its interactions with the MW's own hot coronal gas. The trailing Stream's turbulent morphology seen in \mbox{\HI} is reproduced through interactions between the neutral gas and the warm/hot gas in the Magellanic and Galactic Coronae. We find metal distributions and ionization fractions in agreement with absorption-line spectroscopy observations.}{}

We have \rev{also}{} explored the parameter space of temperatures and densities for the Magellanic Corona to constrain its properties. We find that a \rev{}{pre-infall} mass $>\rev{5}{2}\times10^9$ $\msol$ (within \rev{200}{120} kpc) can provide sufficient ionized material at the present day (Figure~\ref{fig:newbars}) \rev{}{while being stable around the LMC for long times (Figure~\ref{fig:coronagrid})}. \rev{By forming the LMC's gaseous disk self-consistently out of the Corona, we are able to reproduce the gas mass within the galaxy's disk at the present day. The }{In creating our initial conditions for the pre-infall LMC, we found that the} initial temperature of the Coronal gas determines the size and mass of the LMC's \rev{}{gaseous} disk. \rev{ at later times, so we found that a value of $3\times10^5$ K (in agreement with the virial temperature) provides the best results (Figure~\mbox{\ref{fig:lmcdisk}})}{By using a value close to the virial temperature of the LMC's halo, we find that the Coronal gas remains in equilibrium (with the mean temperature closely matching the virial value; Figure~\ref{fig:coronagridtemp}) and the LMC's gaseous disk at the present day agrees well with the observations. This leads to a pre-infall LMC disk gas mass of $9.1\times10^{9}$ $\msol$ with a scale length of 2.8 kpc.}

These models can reproduce the properties of observed Magellanic System while accounting for a large LMC mass. The Magellanic Corona provides the key element that necessitates a review of the precise orbital histories of the Clouds. This brings the Trailing Stream gas much closer to us than previously thought, explaining the turbulent morphology and \Ha\ brightness, and implying an infall onto the MW disk within $\sim$100 Myr.

By obtaining constraints on the distance to the gas in the Stream through absorption-line spectroscopy towards MW halo stars, we will be able to better discriminate between existing models. Moreover, a reevaluation of the properties of the Leading Arm could confirm whether it is Magellanic or not. Further observations like these are needed to constrain key properties of the Magellanic System and to converge on the true history of the Magellanic Clouds.\\[2ex]

\noindent
Support for programs 16363 and 16602 was provided by NASA through a grant from the Space Telescope Science Institute, which is operated by the Association of Universities for Research in Astronomy, Inc., under NASA contract NAS5-26555. The simulations in this paper were run at the University of Wisconsin -- Madison Center for High Throughput Computing supercomputing cluster \citep{uwhpc}.\\

\software{astropy \citep{astropy:2013,astropy:2018,astropy:2022},
dice \citep{dice},
gala \citep{gala,gala_zenodo},
pygad \citep{pygad,rottgers20},
trident \citep{trident}}


\newpage
\bibliographystyle{mnras}
\bibliography{references}

\label{lastpage}

\end{document}